\documentclass[twocolumn,preprintnumbers,superscriptaddress,prd,nofootinbib]{revtex4-1}
\usepackage{graphicx}
\usepackage{times}
\usepackage{epsfig}
\usepackage{amsmath}
\usepackage{amsfonts}
\usepackage{amssymb}
\usepackage{url}
\usepackage{subfigure}
\newcommand{\be}{\begin{equation}}
\newcommand{\ee}{\end{equation}}
\newcommand{\bea}{\begin{eqnarray}}
\newcommand{\eea}{\end{eqnarray}}

\begin{document}

\pagestyle{plain}

\title{$Z^\prime$ portal dark matter in the minimal $B-L$ model}

\author{Satomi Okada}
\affiliation{
Graduate School of Science and Engineering, Yamagata University,
Yamagata 990-8560, Japan
}



\begin{abstract}

In this review article, we consider a dark matter scenario in the context of the minimal extension of the  Standard Model (SM) 
  with a $B-L$ (baryon number minus lepton number) gauge symmetry, 
  where three right-handed neutrinos with a $B-L$ charge $-1$ and a $B-L$ Higgs field with a $B-L$ charge $+2$
  are introduced to make the model anomaly-free and to break the $B-L$ gauge symmetry, respectively. 
The $B-L$ gauge symmetry breaking generates Majorana masses for the right-handed neutrinos. 
We introduce a Z$_2$ symmetry to the model and assign an odd parity only for one right-handed neutrino, 
  and hence the Z$_2$-odd right-handed neutrino is stable and the unique dark matter candidate in the model. 
The so-called minimal seesaw works with the other two right-handed neutrinos 
  and reproduces the current neutrino oscillation data. 
We consider the case that the dark matter particle communicates with the SM particles 
  through the $B-L$ gauge boson ($Z^{\prime}_{B-L}$ boson), 
  and obtain a lower bound on the $B-L$ gauge coupling ($\alpha_{B-L}$) 
  as a function of the $Z^{\prime}_{B-L}$ boson mass ($m_{Z^{\prime}}$) 
  from the observed dark matter relic density. 
On the other hand, we interpret the recent LHC Run-2 results on the search for a $Z^{\prime}$ boson resonance 
   to an upper bound on $\alpha_{B-L}$ as a function of $m_{Z^{\prime}}$. 
These two constraints are complementary to narrow down an allowed parameter region 
   for this “$Z^{\prime}$ portal” dark matter scenario, 
   leading to a lower mass bound of $m_{Z^{\prime}} \ge 3.9$ TeV.

\end{abstract}
\maketitle

\section{Introduction}

In 2012, the Higgs boson, which is the last piece of the Standard Model (SM), was finally discovered 
  by the A Toroidal LHC ApparatuS (ATLAS) and Compact Muon Solenoid (CMS) experiments 
  at the Cern Large Hadron Collider (LHC) \cite{Aad:2012tfa, Chatrchyan:2012xdj}.
The SM is the best theory to describe elementary particles and fundamental interactions among them
   (strong, weak, and electromagnetic interactions), and agrees with a number of experimental results in a high accuracy.
For example, properties of the weak gauge bosons ($W$ and $Z$) in the SM, 
  such as their masses and couplings with the quarks and leptons,
   were measured at the Large Electron-Positron collider (LEP) with a very high degree of precision \cite{LEP:2003, LEP:2013}.
Properties of the Higgs boson have also been measured to be consistent with the SM predictions 
   at the LHC \cite{Khachatryan:2016vau}.

Despite of its great success, there are some observational problems that the SM cannot account for.
One of the major missing pieces of the SM is the neutrino mass matrix.
Since, in contrast to the other fermions, right-handed partners of the SM left-handed neutrinos 
   are missing in the SM particle content, the SM neutrinos cannot acquire their masses 
   at the renormalizable level, even after the electroweak symmetry is broken. 
However, neutrino oscillation phenomena among three neutrino flavors have been confirmed by
   the Super-Kamiokande experiments in 1998 \cite{SuperKamiokande:1998}
   and the Sudbury Neutrino Observatory (SNO) in 2001 \cite{Ahmad:2001an}.
Neutrino oscillation phenomena require neutrino masses and flavor mixings, and 
   therefore we need a framework beyond the SM to incorporate them. 
The so-called type-I seesaw mechanism \cite{Minkowski:1977sc, Yanagida:1979as, Mohapatra:1979ia, Glashow:1979nm, GellMann:1980vs} is a natural way for this purpose, where heavy Majorana right-handed neutrinos are introduced.

Another major missing piece of the SM is a candidate for the dark matter particle in the present universe. 
Based on the recent results of the precision measurements of the cosmic microwave background (CMB) anisotropy 
   by the Wilkinson Microwave Anisotropy Probe (WMAP) \cite{Hinshaw:2012aka} and the Planck satellite \cite{Ade:2013kta, Ade:2015xua},
   the energy budget of the present universe is determined 
   to be composed of 73\% dark energy, 23\% cold dark matter and only 4\% from baryonic matter.
Since the SM has no suitable candidate for the (cold) dark matter particle, 
   we need to extend the SM to incorporate it. 
The weakly interacting massive particle (WIMP) \cite{Lee:1977ua} has long been studied 
   as one of the most promising candidates for the dark matter. 
Through its interaction with the SM particles, the WIMP was in thermal equilibrium in the early universe 
   and the WIMP dark matter is a thermal relic from the early universe. 
Note that the relic density of the WIMP dark matter is independent of the history of the universe 
   before it has gotten in thermal equilibrium.

Among many possibilities, 
   the minimal gauged $B-L$ extension of the SM \cite{Mohapatra:1980qe, Marshak:1979fm, Wetterich:1981bx, Masiero:1982fi, Buchmuller:1991ce}
   is a very simple way to incorporate neutrino masses and flavor mixings via the seesaw mechanism. 
In this extension, the accidental global $B-L$ (baryon number minus lepton number) symmetry in the SM is gauged. 
Associated with this gauging, three right-handed neutrinos with a $B-L$ charge $-1$ are introduced to cancel 
   all the gauge and mixed-gravitational anomalies of the model. 
In other words, the right-handed neutrinos which play the essential role in the type-I seesaw mechanism 
   must present for the theoretical consistency. 
A SM gauge singlet Higgs boson with a $B-L$ charge $-2$ is also contained in the model and its vacuum expectation value (VEV) 
   breaks the $B-L$ gauge symmetry. 
The Higgs VEV generates the $B-L$ gauge boson mass as well as Majorana masses of the right-handed neutrinos. 
After the electroweak symmetry breaking, the SM neutrino Majorana masses are generated through the seesaw mechanism. 
The mass spectrum of new particles introduced in the minimal $B-L$ model, 
  the $B-L$ gauge boson ($Z_{B-L}^{\prime}$ boson), the right-handed Majorana neutrinos, 
  and the $B-L$ Higgs boson, is controlled by the $B-L$ symmetry breaking scale. 
If the breaking scale lies around the TeV scale the minimal $B-L$ model can be tested at the LHC in the future.

Although the minimal $B-L$ model supplements the SM with the neutrino masses and mixings, 
   a cold dark matter candidate is still missing. 
Towards a more complete scenario, we need to consider a further extension of the model. 
Ref.~\cite{Okada:2010wd} has proposed a concise way to introduce a dark matter candidate to the minimal $B-L$ model, 
   where instead of introducing a new particle as a dark matter candidate, a Z$_2$ symmetry is introduced 
   and an odd parity is assigned only for one right-handed neutrino. 
Thanks to the Z$_2$ symmetry,  the Z$_2$-odd right-handed neutrino becomes stable and hence plays the role of dark matter. 
On the other hand, the other two right-handed neutrinos are involved in the seesaw mechanism. 
It is known that the type-I seesaw with two right-handed neutrinos is a minimal system 
   to reproduce the observed neutrino oscillation data. 
This so-called minimal seesaw \cite{King:1999mb, Frampton:2002qc} predicts one massless neutrino.
Dark matter phenomenology in this model context has been investigated in Refs.~\cite{Okada:2010wd, Okada:2012sg, Basak:2013cga}.
The right-handed neutrino dark matter can communicate with the SM particles through 
 (i) the $Z_{B-L}^{\prime}$ boson and (ii) two Higgs bosons which are realized as linear combinations of the SM Higgs and
   the $B-L$ Higgs bosons.
The case (i) and (ii) are, respectively, called “$Z^{\prime}$ portal” and “Higgs portal” dark matter scenarios. 
In the following, we focus on the “$Z^{\prime}$ portal” dark matter scenario. 
See Refs.~ \cite{Okada:2010wd, Okada:2012sg, Basak:2013cga} for extensive studies 
   on the Higgs portal dark matter scenario.

In recent years, the $Z^{\prime}$ portal dark matter has attracted a lot of attention
   \cite{          
An:2012va, An:2012ue,  Soper:2014ska,
Burell:2011wh, Basso:2012ti, Das:2013jca, Chu:2013jja, Dudas:2013sia, Lindner:2013awa, Alves:2013tqa, Kopp:2014tsa, 
Agrawal:2014ufa,Hooper:2014fda,Ma:2014qra, Alves:2015pea,  Ghorbani:2015baa, Sanchez-Vega:2015qva, Duerr:2015wfa,Ma:2015mjd, 
Okada:2016gsh, Okada:2016tzi, Chao:2016avy, Biswas:2016ewm, Fairbairn:2016iuf, 
Klasen:2016qux, Dev:2016xcp, Altmannshofer:2016jzy, Okada:2016tci, Kaneta:2016vkq, Alves:2016cqf, 
Singirala:2017cch,
Arcadi:2017kky, 
Cui:2017juz, 
Arcadi:2017jqd,
Bandyopadhyay:2017bgh,
DeRomeri:2017oxa,Nanda:2017bmi, Profumo:2017obk},
   where a dark matter candidate along with a new $U(1)$ gauge symmetry is introduced 
   and the dark matter particle communicates with the SM particles through the $U(1)$ gauge boson 
   ($Z^{\prime}$ boson).      
Through this $Z^\prime$ boson interaction, we can investigate a variety of dark matter physics,
   such as the dark matter relic density and the direct/indirect dark matter search. 
Very interestingly, the search for a $Z^{\prime}$ boson production by the LHC experiments 
   can provides the information which is complementary to the information obtained by dark matter physics.

Note that the minimal $B-L$ model with the right-handed neutrino dark matter introduced above 
   is a simple example of the $Z^{\prime}$ portal dark matter scenario. 
In this article, we consider this $Z^{\prime}$ portal dark matter scenario. 
Since the model is very simple, dark matter physics is controlled by only three free parameters, 
   namely, the $B-L$ gauge coupling ($\alpha_{B-L}$), the $Z_{B-L}^{\prime}$ boson mass ($m_{Z^{\prime}}$), 
   and the dark matter mass ($m_{\textrm{DM}}$).
We first identify allowed parameter regions of the model by considering the cosmological bound on 
   the dark matter relic density.  
We then consider the results from the search for a $Z^\prime$ boson resonance with dilepton final states 
   to identify allowed parameter regions. 
Combining the cosmological and the LHC constraints, we find a narrow allowed region. 
This complementary between the cosmological and the LHC constraints has been investigated 
   in Refs.~\cite{Okada:2016gsh, Okada:2016tci}. 
The purpose of this article is to update the results in the references by employing the latest LHC results, 
   along with a review of the minimal $B-L$ model with the right-handed neutrino dark matter.

The plan of this article is as follows: 
In the next section, we give a review of the minimal $B-L$ model. 
In Sec.~\ref{sec:3}, we introduce the minimal $B-L$ model with $Z_2$ symmetry, 
  where one right-handed neutrino, which is a unique $Z_2$-odd particle in the model,  
  is identified with the dark matter particle. 
In  Sec.~\ref{sec:4}, the cosmological constraints on the right-handed neutrino dark matter 
  is considered and an allowed parameter region is identified. 
In Sec.~\ref{sec:5}, we consider the LHC Run-2 constraints from the search for a narrow resonance with dilepton final states  
  and find the constraints on our model parameters. 
Combining the results obtained in Sec.~\ref{sec:3}, we find an allowed region. 
The cosmological constraint and the LHC constraints are complementary to narrow down 
  the allowed parameter regions. 
Sec.~\ref{sec:6} is devoted to conclusions.

\section{The minimal $B-L$ model}
%
\begin{table}[t]
\begin{center}
  \begin{tabular}{c||ccc|c}
    & $SU(3)_C$ & $SU(2)_L$ & $U(1)_Y$ & $U(1)_{B-L}$ \\ \hline
    $q_L^i$ & $\bf{3}$ & $\bf{2}$ & $1/6$ & $1/3$ \\
    $u_R^i$ & $\bf{3}$ & $\bf{1}$ & $2/3$ & $1/3$ \\
    $d_R^i$ & $\bf{3}$ & $\bf{1}$ & $-1/3$ & $1/3$ \\ \hline
    $l_L^i$ & $\bf{1}$ & $\bf{2}$ & $-1/2$ & $-1$ \\
    $N_R^i$ & $\bf{1}$ & $\bf{1}$ & $0$ & $-1$ \\
    $e_R^i$ & $\bf{1}$ & $\bf{1}$ & $-1$ & $-1$ \\ \hline
    $H$ & $\bf{1}$ & $\bf{2}$ & $-1/2$ & $0$ \\
    $\Phi$ & $\bf{1}$ & $\bf{1}$ & $0$ & $2$
  \end{tabular}
  \renewcommand{\baselinestretch}{1.1}
  \caption{Particle content of the minimal $B-L$ model.} \label{BLcontent}
  \end{center}
\end{table}

The SM Lagrangian at the tree-level is invariant under the global $U(1)_B$ and $U(1)_L$ transformations,
\begin{eqnarray}
 \psi &\to& \psi' = {\rm{e}}^{iQ_B\theta_B} \psi, \nonumber \\
 \psi &\to& \psi' = {\rm{e}}^{iQ_L\theta_L} \psi,
\end{eqnarray}
   where $\theta_B$ and $\theta_L$ are constant phases associated with the $U(1)_B$ and $U(1)_L$ transformations,
   and $Q_B$ and $Q_L$ are charges identified as a baryon number ($B$) and a lepton number ($L$) of the fermion $\psi$, respectively.
The baryon number is a quantum number to characterize fermions.
A quark (antiquark) has a baryon number $1/3$ ($-1/3$), while a SM lepton has 0.
The lepton number is a quantum number similar to baryon number.
A lepton (antilepton) has a lepton number 1 ($-1$), while a quark has 0.
Although these $U(1)$ symmetries are anomalous under the SM gauge group, the combination of $B-L$ is anomaly free.
The $B-L$ symmetry means that the SM Lagrangian is invariant under the global $U(1)_{B-L}$ transformation, 
\begin{eqnarray}
 \psi \ \to \ \psi' = {\rm{e}}^{i(Q_B-Q_L)\theta_{B-L}} \psi,
\end{eqnarray}
   where $\theta_{B-L}$ is a constant phase associated with the $U(1)_{B-L}$ transformation.

In the minimal $B-L$ model \cite{Mohapatra:1980qe, Marshak:1979fm, Wetterich:1981bx, Masiero:1982fi, Buchmuller:1991ce},
   this global $B-L$ symmetry in the SM is gauged,
   and hence this model is based on the gauge group $SU(3)_C \times SU(2)_L \times U(1)_Y \times U(1)_{B-L}$.
Three right-handed neutrinos ($N_R^i$,  $i=1,2,3$ is a generation index) and an SM singlet scalar field ($\Phi$) are introduced
   to make the theory anomaly free, and to break the $U(1)_{B-L}$ gauge symmetry, respectively.
The particle content of the minimal $B-L$ model is listed in Table \ref{BLcontent}.

\subsection{Gauge sector}

Lagrangian of the gauge bosons in the $B-L$ model is generally given by
\begin{eqnarray}
 {\cal{L}}_{\textrm{gauge}}^{B-L}
  =&& - \frac{1}{2}{\rm{tr}}[G_{\mu\nu}G^{\mu\nu}] - \frac{1}{2}{\rm{tr}}[F_{\mu\nu}F^{\mu\nu}] \nonumber \\
    && - \frac{1}{4}B_{\mu\nu}B^{\mu\nu} - \frac{1}{4}B'_{\mu\nu}B'^{\mu\nu} - c_{\textrm{mix}} B_{\mu\nu}B'^{\mu\nu},
\end{eqnarray}
   where $G_{\mu \nu}$, $F_{\mu \nu}$ and $B_{\mu \nu}$ are the field strengths
   of the SM gauge fields of $SU(3)_C \times SU(2)_L \times U(1)_Y$, 
   and    
\begin{eqnarray}
 B'_{\mu\nu} = \partial_{\mu} (Z'_{B-L})_{\nu} - \partial_{\nu} (Z'_{B-L})_{\mu}
\end{eqnarray}
   is the field strength for the new electrically neutral gauge boson ($Z'_{B-L}$) of $U(1)_{B-L}$.
Note that we can generally introduce the last term for a kinetic mixing between the $U(1)_Y$ and the $U(1)_{B-L}$ gauge bosons.
In fact, such a mixing term is generated through quantum corrections.  
Since we can always set $c_{\textrm{mix}} = 0$ at a fixed energy, 
  we define the minimal $B-L$ model with $c_{\textrm{mix}} = 0$ 
  at the scale of the $B-L$ symmetry breaking, for simplicity.

\subsection{Scalar sector}

Lagrangian of the scalar sector in the minimal $B-L$ model is given by
\begin{eqnarray}
 {\cal{L}}_{\textrm{scalar}}^{B-L}
  =&& ({\cal{D}}^{\mu}H)^{\dagger} ({\cal{D}}_{\mu}H)
         + ({\cal{D}}^{\mu}\Phi)^{\dagger} ({\cal{D}}_{\mu}\Phi) \nonumber \\
    && - V(H,\Phi),
\end{eqnarray}
   where $H$ and $\Phi$ are the SM Higgs field and the SM singlet scalar field ($B-L$ Higgs), respectively,
   and the scalar potential is given by
\begin{eqnarray} \label{BLpoten}
  V(H,\Phi)
   &=& \lambda_H \left( H^{\dagger}H - \frac{v^2}{2} \right)^2
   + \lambda_{\Phi} \left( \Phi^{\dagger}\Phi - \frac{v^2_{B-L}}{2} \right)^2 \nonumber\\
   && + \lambda_{\textrm{mix}} \left( H^{\dagger}H - \frac{v^2}{2} \right) \left( \Phi^{\dagger}\Phi - \frac{v^2_{B-L}}{2} \right).
\end{eqnarray}
Here, $\lambda_H \ (>0)$, $\lambda_{\Phi} \ (>0)$ and $\lambda_{\textrm{mix}}$ are real coupling constants,
   $v = 246$ GeV \cite{PDG:2016}, and $v_{B-L}$ is a real and positive constant.
We will derive a condition for $\lambda_{\textrm{mix}}$ to make the potential bounded from below.
In this scalar potential, the SM Higgs doublet and $U(1)_{B-L}$ Higgs field develop the VEVs,
\begin{eqnarray}
\langle H \rangle
 &=&
\frac{1}{\sqrt{2}}
\left(
 \begin{array}{c}
 v\\
 0
 \end{array}
\right), \nonumber\\
\langle \Phi \rangle
 &=&
\frac{v_{B-L}}{\sqrt{2}}.
\end{eqnarray}
We expand the Higgs fields around the VEVs such that
\begin{eqnarray}
H &=&
\frac{1}{\sqrt{2}}
\left(
 \begin{array}{c}
 v + h\\
 0
 \end{array}
\right), \nonumber\\
\Phi &=&
\frac{v_{B-L}+h'}{\sqrt{2}},
\end{eqnarray}
   where $h$, $h'$ are physical Higgs bosons.
Substituting this expansion into the scalar potential (\ref{BLpoten}), we read out the mass terms of the Higgs bosons as
\begin{eqnarray}
 V(H,\Phi)
 &\supset&
 \lambda_H v^2 h^2 + \lambda_{\Phi} v_{B-L}^2 h'^2 + \lambda_{\textrm{mix}} v v_{B-L} h h' \nonumber\\
 &=&
 \frac{1}{2} (h \ h')
 \left(
 \begin{array}{cc}
 2 \lambda_H v^2                                & \lambda_{\textrm{mix}} v v_{B-L} \\
 \lambda_{\textrm{mix}} v v_{B-L} & 2 \lambda_{\Phi} v_{B-L}^2
 \end{array}
 \right)
 \left(
 \begin{array}{c}
 h \\
 h'
 \end{array}
 \right) \nonumber \\
 &=& \frac{1}{2} (h \ h') M_{\textrm{scalar}}
     \left(
 \begin{array}{c}
 h \\
 h'
 \end{array}
 \right).
\end{eqnarray}
In order for the scalar potential to be bounded from below, the mass matrix $M_{\textrm{scalar}}$ must be positive definite, namely,
\begin{eqnarray}
 {\rm det} [ M_{\textrm{scalar}} ]
 = ( 4 \lambda_H \lambda_{\Phi} - \lambda_{\textrm{mix}}^2 ) v^2 v_{B-L}^2 > 0,
\end{eqnarray}
   and hence $|\lambda_{mix}| < 2 \sqrt{\lambda_H \lambda_{\Phi}}$.
Now we diagonalize the mass matrix by
\begin{eqnarray} \label{BLHmass}
 \left(
 \begin{array}{c}
 h \\
 h'
 \end{array}
 \right)
 =
  \left(
 \begin{array}{cc}
 \cos{\alpha} & \sin{\alpha} \\
 -\sin{\alpha} & \cos{\alpha}
 \end{array}
 \right)
 \left(
 \begin{array}{c}
 h_1 \\
 h_2
 \end{array}
 \right),
\end{eqnarray}
   where $h_1$, $h_2$ are mass eigenstates, and the mixing angle is given by
\begin{eqnarray} \label{BLmixingangle}
 \tan{2\alpha} = - \frac{\lambda_{\textrm{mix}} v v_{B-L}}{\lambda_{H} v^2 - \lambda_{\Phi} v_{B-L}^2}.
\end{eqnarray}
The mass eigenstates are given by
\begin{eqnarray}
 m_{h_1}^2
 = \ \ 
 && \lambda_H v^2 + \lambda_{\Phi} v_{B-L}^2 \nonumber \\
 && + \sqrt{(\lambda_H v^2 - \lambda_{\Phi} v_{B-L}^2)^2 + (\lambda_{\textrm{mix}} v v_{B-L})^2}, \nonumber \\
 m_{h_2}^2
 = \ \ 
 && \lambda_H v^2 + \lambda_{\Phi} v_{B-L}^2 \nonumber \\
 && - \sqrt{(\lambda_H v^2 - \lambda_{\Phi} v_{B-L}^2)^2 + (\lambda_{\textrm{mix}} v v_{B-L})^2}.
\end{eqnarray}
For simplicity, we assume a very small $\lambda_{\textrm{mix}}$, so that one mass eigenstate is an SM-like Higgs boson,
   and the other is almost a $B-L$ Higgs boson.
Since the Higgs boson properties measured by the LHC experiments are consistent with the SM predictions \cite{Khachatryan:2016vau}, 
   $|\lambda_{\textrm{mix}}| \ll1$ is justified.

Let us now calculate the mass of the $B-L$ gauge boson $Z^{\prime}_{B-L}$.
The kinetic term of the $B-L$ Higgs field is given by
\begin{eqnarray}
 {\cal{L}}^{B-L \ \textrm{kin}}_{\textrm{scalar}} = ({\cal{D}}_{\mu} \Phi)^{\dagger} ({\cal{D}}^{\mu} \Phi),
\end{eqnarray}
   where the covariant derivative is
\begin{eqnarray}
 {\cal{D}}_{\mu} = \partial_{\mu} - 2ig_{B-L} (Z^{\prime}_{B-L})_{\mu},
\end{eqnarray}
   and $g_{B-L}$ is the coupling constant of $U(1)_{B-L}$ gauge interaction.
Substituting $\Phi \to \langle \Phi \rangle$, the $Z'_{B-L}$ gauge boson mass is found to be
\begin{eqnarray}
 m_{Z'} = 2g_{B-L} v_{B-L}.
\end{eqnarray}

\subsection{Yukawa sector}

Lagrangian of the Yukawa sector in the $B-L$ model is given by
\begin{eqnarray}
 {\cal{L}}_{\textrm{Yukawa}}^{B-L}
 \supset \ 
 && - \sum^3_{i=1} \sum^3_{j=1} Y^{ij}_D \overline{l^i_L} H N^j_R
       - \frac{1}{2} \sum^3_{k=1} Y^k_N \Phi \overline{N^{kC}_R} N^k_R \nonumber \\
 && + H.c.,
 \label{Yukawa_BL}
\end{eqnarray}
   where $Y_D^{ij}$ and $Y_N^k$ are Dirac Yukawa coupling constant and Majorana Yukawa coupling constant.
Once the $B-L$ Higgs field $\Phi$ develops its VEV,
   the $B-L$ gauge symmetry is broken and the Majorana mass terms for the right-handed neutrinos are generated
   from the second term in the right-hand side.
The seesaw mechanism is automatically implemented in the model after the electroweak symmetry breaking.
Neutrino mass matrix is given by
\begin{eqnarray}
 M_{\textrm{neutrino}} =
 \left(
  \begin{array}{cc}
    0      & m_D\\
    m_D^T & M
  \end{array} 
 \right).
\end{eqnarray}
Here, $m_D$ and $M$ are Dirac and Majorana mass matrices, respectively, which are given by
\begin{eqnarray}
 m_D &=& \frac{Y_D}{\sqrt{2}} v, \nonumber \\
 M      &=& \frac{Y_N}{\sqrt{2}} v_{B-L}.
\end{eqnarray}
Assuming $|m_D^{ij}| \ll M^k$, we can block-diagonalize the mass matrix $M_{\textrm{neutrino}}$ to be
\begin{eqnarray}
 \left(
  \begin{array}{cc}
    0          & m_D\\
    m_D^T & M
  \end{array} 
 \right)
 \ \ \to \ \
 \left(
  \begin{array}{cc}
    - m_D^T M^{-1} m_D & 0\\
    0                                & M
  \end{array} 
 \right).
\end{eqnarray}
When we consider only one generation, the mass eigenvalues are simply
\begin{eqnarray}
 m_{\nu_l} &\simeq& - \frac{m_D^2}{M}, \nonumber \\
 m_{\nu_h} &\simeq& M.
\end{eqnarray}
Because of the seesaw mechanism, a huge mass hierarchy between the light eigenstate ($\nu_l$)
   and the heavy eigenstate ($\nu_h$) is generated.

\section{The minimal  $B-L$ model with Z$_2$ parity}
\label{sec:3}

\begin{table}[t]
\begin{center}
  \begin{tabular}{c|ccc|c|c}
    & $SU(3)_C$ & $SU(2)_L$ & $U(1)_Y$ & $U(1)_{B-L}$  & Z$_2$ \\ \hline
    $q_L^i$ & \bf{3} & \bf{2} & 1/6 & 1/3 & + \\
    $u_R^i$ & \bf{3} & \bf{1} & 2/3 & 1/3 & + \\
    $d_R^i$ & \bf{3} & \bf{1} & -1/3 & 1/3 & + \\ \hline
    $l_L^i$ & \bf{1} & \bf{2} & -1/2 & -1 & + \\
    $N_R^j$ & \bf{1} & \bf{1} & 0 & -1 & + \\
    $N_R$ & \bf{1} & \bf{1} & 0 & -1 & - \\
    $e_R^i$ & \bf{1} & \bf{1} & -1 & -1 & + \\ \hline
    $H$ & \bf{1} & \bf{2} & -1/2 & 0 & + \\
    $\Phi$ & \bf{1} & \bf{1} & 0 & 2 & +
  \end{tabular}
  \renewcommand{\baselinestretch}{1.1}
  \caption{
  The particle content of the minimal $B-L$ extended SM with Z$_2$ symmetry.
In addition to the SM particle content ($i = 1$, $2$, $3$), 
  the three right-handed neutrinos [$N_R^j$ ($j = 1$, $2$) and $N_R$] and
  the $B-L$ Higgs field ($\Phi$) are introduced.
Because of the Z$_2$ parity assignment shown here, the $N_R$ is a unique (cold) dark matter candidate.
} 
\label{BLDM}
  \end{center}
\end{table}

In the previous section, we have discussed that the minimal $B-L$ extended SM 
   incorporates the neutrino masses and mixings through the seesaw mechanism. 
In this section, we extend the model further to introduce a cold dark matter candidate in the model. 
Among many possibilities, we follow a very concise way proposed in Ref.~ \cite{Okada:2010wd} 
   and introduce a Z$_2$ symmetry without extending the model particle content. 
We then assign an odd parity only for one right-handed neutrino $N_R$. 
The particle content listed on Table \ref{BLDM}.
Except for the introduction of the Z$_2$ symmetry and the parity assignments, 
   the particle contents is identical to that of the minimal $B-L$ model in Table \ref{BLcontent}. 
The conservation of the Z$_2$ parity ensures the stability of the Z$_2$-odd $N_R$,
   and therefore, this right-handed neutrino is a unique dark matter candidate in the model \cite{Okada:2010wd}.

With the Z$_2$ symmetry, the Yukawa sector of the minimal $B-L$ model  in (\ref{Yukawa_BL}) 
   is modified to be
\begin{eqnarray}
 {\cal{L}}_{\textrm{Yukawa}}^{B-L}
 \supset &&
      - \sum^3_{i=1} \sum^2_{j=1} Y^{ij}_D \overline{l^i_L} H N^j_R
      - \frac{1}{2} \sum^2_{k=1} Y^k_N \Phi \overline{N^{kC}_R} N^k_R \nonumber \\
 && - \frac{1}{2}  Y_N \Phi \overline{N^C_R} N_R + H.c.
\end{eqnarray}
Note that due to the Z$_2$ parity assignment only the two generation right-handed neutrinos 
   are involved in the neutrino Dirac Yukawa coupling.
The renormalizable scalar potential for the SM Higgs and the $B-L$ Higgs fields are the same as the minimal $B-L$ model,
   and the Higgs fields develop their VEVs. 
This $B-L$ symmetry breaking generates masses for 
   the Majorana neutrinos $N^j_R$ ($j=1$, $2$), the dark matter particle $N_R$
   and the $B-L$ gauge boson ($Z_{B-L}^{\prime}$ boson):
\begin{eqnarray}
 m^j_N    &=& \frac{Y^j_N}{\sqrt{2}} v_{B-L}, \nonumber \\
 m_{\textrm{DM}} &=& \frac{Y_N}{\sqrt{2}} v_{B-L}, \nonumber \\
 m_{Z'}    &= &2 g_{B-L} v_{B-L}.
\end{eqnarray}
The seesaw mechanism \cite{Minkowski:1977sc, Yanagida:1979as, Mohapatra:1979ia, Glashow:1979nm, GellMann:1980vs}
   is automatically implemented in the model after the electroweak symmetry breaking.
Due to the Z$_2$ symmetry, 
    only two right-handed neutrinos $N_R^{1, 2}$ are relevant to the seesaw mechanism. 
This system is the so-called minimal seesaw \cite{King:1999mb, Frampton:2002qc}
    which possesses a number of free parameters
   $Y^{ij}_D$ and $Y^k_N$ enough to reproduce the neutrino oscillation data with predicting 
   one massless eigenstate. 
Since the lightest neutrino is massless in our model, 
 the pattern of the light neutrino mass spectrum is either the normal hierarchy or the inverted hierarchy. 
The quasi-degenerate mass spectrum cannot be realized.

The dark matter particle can communicate with the SM particles in two ways: 
One is through the Higgs bosons. 
In the Higgs potential of (\ref{BLpoten}),
   the SM Higgs boson and the $B-L$ Higgs boson mix with each other in the mass eigenstates (see (\ref{BLHmass}) and (\ref{BLmixingangle})),
   and this Higgs boson mass eigenstates mediate the interactions between the dark matter particle and the SM particles.
Dark matter physics with the Higgs interactions have been investigated in Refs.~\cite{Okada:2010wd, Okada:2012sg, Basak:2013cga}.
In this analysis,  four free parameters are involved, namely, 
   the dark matter mass, Yukawa coupling $Y_N$, the $B-L$ Higgs boson mass, and 
   a mixing parameter between the SM Higgs and $B-L$ Higgs bosons.  
The other way for the dark matter particle to communicate with the SM particles
   is through the $B-L$ gauge interaction with the $Z'_{B-L}$ gauge boson.
In this case, only three free parameters ($g_{B-L}$, $m_{Z'}$ and $m_{DM}$) are involved in dark matter physics analysis.
In this article, we concentrate on dark matter physics mediated by the $Z_{B-L}^{\prime}$ boson, namely ``$Z^{\prime}$ portal dark matter.''
Assuming $|\lambda_{\textrm{mix}}| \ll 1$ in the Higgs potential (\ref{BLpoten}), 
   the Higgs bosons mediated interactions are negligibly small, 
   and the dark matter particle communicates with the SM particles only through the $Z_{B-L}^{\prime}$ boson. 
We may consider a supersymmetric extension of our model \cite{Burell:2011wh} 
   to naturally realize this situation, where $\lambda_{\textrm{mix}}$ is forbidden by supersymmetry. 
Although we do not consider the supersymmetric case, 
   dark matter phenomenology in our model is essentially the same as the supersymmetric case (see Ref.~\cite{Burell:2011wh}).
   when all the superpartners of the SM particles are heavier than the dark matter particle.    
See Refs.~\cite{Okada:2012sg, Basak:2013cga, Burell:2011wh, Oda:2017kwl} for studies on 
    the $Z_{B-L}^{\prime}$ portal dark matter scenario with a limited parameter choice.

\section{Cosmological constraints on $Z_{B-L}^{\prime}$ portal dark matter}
\label{sec:4}

%
\begin{figure}[tb]
 \centering
  \includegraphics[width=30mm, angle=90]{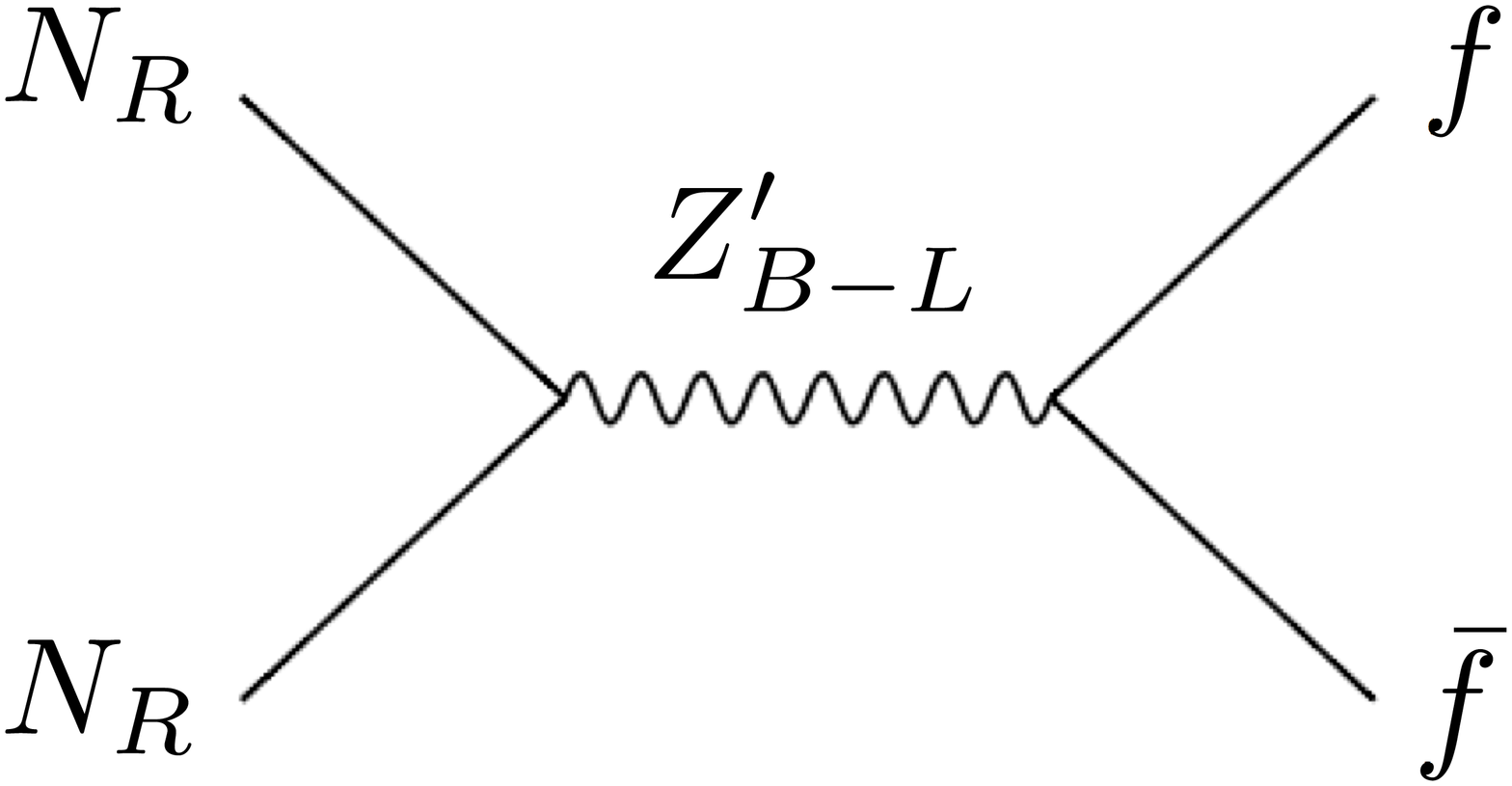}
  \includegraphics[width=30mm, angle=90]{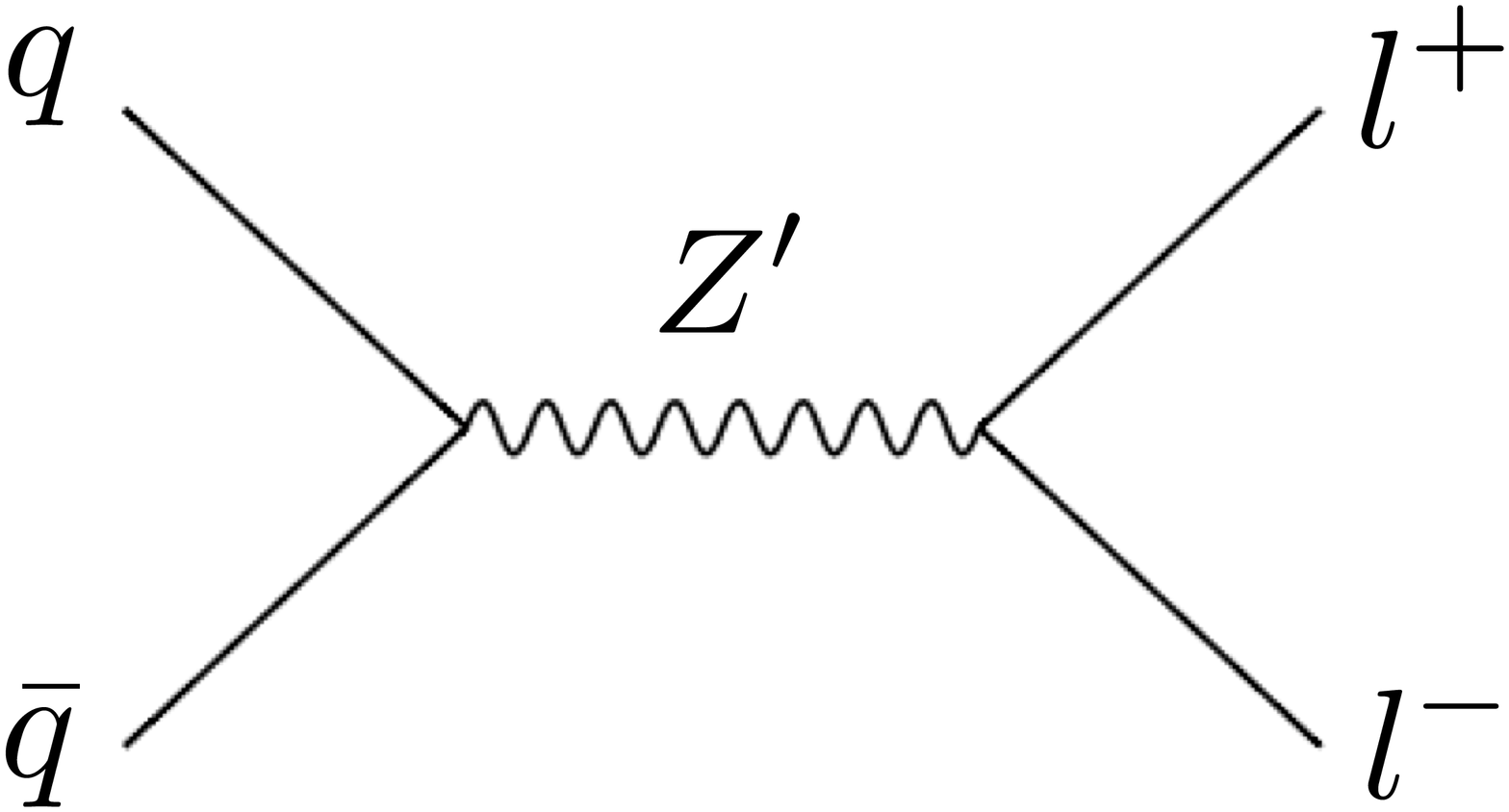}
 \renewcommand{\baselinestretch}{1.1}
 \caption{Left:
 Majorana neutrino dark matter ($N_R$) pair annihilation process into the SM fermions ($f$) through the $Z'_{B-L}$ exchange in the $s$-channel, 
 $N_R N_R \to Z'_{B-L} \to f\bar{f}$.
 Right:
 parton level process (quark ($q$) and anti-quark ($\bar{q}$) annihilation process) to produce a dilepton final state ($l^+ l^-$)
 through a $Z^\prime$ exchange in the $s$-channel at the LHC.
 } 
  \label{BLcomplementarity}
\end{figure}
%

The dark matter relic density is measured at the 68$\%$ limit as \cite{Aghanim:2015xee}
\begin{eqnarray} \label{cosmologicalconstraint}
 \Omega_{\textrm{DM}} h^2
 = 0.1198 \pm 0.0015.
\end{eqnarray}
We now evaluate the relic density of the dark matter $N_R$ and identify an allowed parameter region that satisfies the upper bound on the dark matter relic density of $\Omega_{\textrm{DM}} h^2 \le 0.1213$.
The relic density of dark matter $N_R$ is evaluated by solving the Boltzmann equation:
\begin{eqnarray} \label{BLBoltzmanE}
  \frac{dY_{\textrm{DM}}}{dx}
 =
 - \frac{s \langle \sigma v_{\textrm{rel}} \rangle}{xH(m_{\textrm{DM}})} (Y_{\textrm{DM}}^2 - (Y_{\textrm{DM}}^{\textrm{eq}})^2)
\end{eqnarray}
where $Y_{\textrm{DM}} = n_{DM}/s$ is the yield of the dark matter particle with the dark matter number density ($n_{DM}$)
and the entropy density ($s$),
$Y_{DM}$ in thermal equilibrium is denoted as $Y^{\textrm{eq}}_{\textrm{DM}}$,
$x \equiv m_{\textrm{DM}}/T$ ($T$ is temperature of the universe) is time normalized by the dark matter mass,
$H(m_{\textrm{DM}})$ is the Hubble parameter at $T=m_{\textrm{DM}}$,
and $\langle \sigma v_{\textrm{rel}} \rangle$ is the thermal average of the cross section for dark matter annihilation process times relative velocity.
We give explicit formulas of the quantities in the Boltzmann equation:
\begin{eqnarray}
 s &=& \frac{2\pi^2}{45} g_* \frac{m_{\textrm{DM}}^3}{x^3}, \nonumber \\
 H(m_{\textrm{DM}}) &=& \sqrt{\frac{4\pi^3}{45} g_*} \frac{m_{\textrm{DM}}^2}{M_{pl}}, \nonumber \\
 s Y^{\textrm{eq}}_{\textrm{DM}} &=& \frac{g_{\textrm{DM}}}{2\pi^2} \frac{m_{\textrm{DM}}^3}{x} K_2(x),
\end{eqnarray}
where $M_{pl} = 1.22 \times 10^{19}$ GeV is the Planck mass,
 $g_*$ is the effective total degrees of freedom for SM particles in thermal equilibrium
 ($g_* = 106.75$ is employed in the following analysis), 
 $g_{\textrm{DM}} = 2$ is the degrees of freedom for the right-handed neutrino dark matter,
  and $K_2$ is the modified Bessel function of the second kind.
In our $Z'_{B-L}$ portal dark matter scenario, 
  the dark matter particles pair-annihilate into the SM particles mainly 
  through the $s$-channel $Z'_{B-L}$ boson exchange (see the left panel of Figure \ref{BLcomplementarity}).
The thermal average of the annihilation cross section is calculated as 
\begin{eqnarray} \label{avesv}
 \langle \sigma v_{\textrm{rel}} \rangle
 = &&
      (sY_{\textrm{DM}}^{\textrm{eq}})^{-2} g_{\textrm{DM}}^2 \frac{m_{\textrm{DM}}}{64\pi^4 x} \nonumber \\
 &&\times \int^{\infty}_{4m_{\textrm{DM}}^2} ds \hat{\sigma}(s) \sqrt{s} K_1 \left( \frac{x \sqrt{s}}{m_{\textrm{DM}}} \right),
\end{eqnarray}
where $\hat{\sigma}(s) = 2(s-4m_{\textrm{DM}}^2) \sigma(s)$ is the reduced cross section 
  with $\sigma(s)$ being the total annihilation cross section. 
The total cross section of the annihilation process $N_R N_R \to Z'_{B-L} \to f\bar{f}$ ($f$ denotes an SM fermion)
  is calculated as
\begin{eqnarray} \label{xsection}
 \sigma(s)
 = &&
 \pi \alpha_{B-L}^2 \frac{\sqrt{s( s - 4m_{\textrm{DM}}^2 )}}{( s - m_{Z'}^2 )^2 + m_{Z'}^2 \Gamma_{Z'}^2} \nonumber \\
 &&\times \left[
 \frac{37}{9} + \frac{1}{3} \beta_t \left( 1 - \frac{1}{3} \beta_t^2 \right)
 \right]
 \label{annihilation_X}
\end{eqnarray}
with $\beta_t(s) = \sqrt{1 - 4m_t^2/s}$, top quark mass of $m_t = 173.34$ GeV \cite{PDG:2016}
 and the total decay width of $Z'_{B-L}$ boson given by
\begin{eqnarray}
 \Gamma_{Z'}
 = &&
 \frac{\alpha_{B-L}}{6} m_{Z'}
 \left[
 \frac{37}{3} + \frac{1}{3} \beta_t (m_{Z'}^2) (3 - \beta_t (m_{Z'}^2)^2) \right. \nonumber \\
 &&
 \left. + \left( 1 - \frac{4m_{\textrm{DM}}^2}{m_{Z'}^2} \right)^{\frac{2}{3}}
 \theta \left( \frac{m_{Z'}^2}{m_{\textrm{DM}}^2} - 4 \right)
 \right]
\end{eqnarray}
Here, we have taken $m^j_N > m_{Z'}/2$, for simplicity.

\begin{figure}[tb]
 \begin{center}
  \includegraphics[width=80mm]{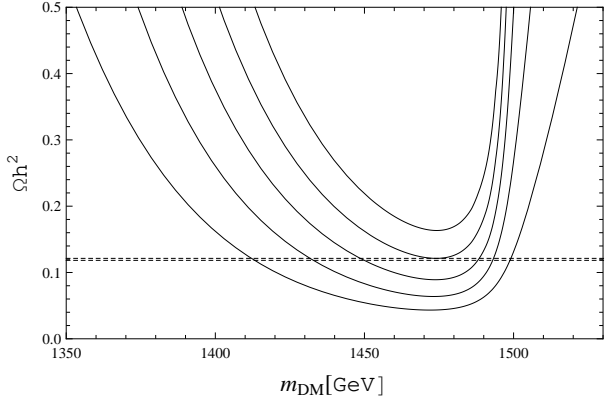}
 \end{center}
 \renewcommand{\baselinestretch}{1.1}
 \caption{The relic abundance of the $Z'_{B-L}$ portal right-hard neutrino dark matter
 as a function of the dark matter mass ($m_{\textrm{DM}}$) for $m_{Z'} = 3$ TeV and various values of the gauge coupling
 $\alpha_{B-L} = 0.001, \ 0.0014, \ 0.002, \ 0.003$ and $0.005$ (solid lines from top to bottom).
 The two horizontal lines denote the range of the observed dark matter relic density,
 $0.1183 \le \Omega_{\textrm{DM}} h^2 \le 0.1213$.} \label{BLomega}
\end{figure}

Solving the Boltzmann equation numerically, we evaluate the dark matter relic density by 
\begin{eqnarray}
 \Omega_{\textrm{DM}} h^2
 = \frac{m_{\textrm{DM}} s_0 Y(\infty)}{\rho_{crit}/h^2},
\end{eqnarray}
where $Y(\infty)$ is the yield in the limit of $x \to \infty$, 
  $s_0 = 2890$ cm$^{-3}$ is the entropy density of the present universe,
  and $\rho_{\textrm{crit}}/h^2 = 1.05 \times 10^{-5}$ GeV/cm$^3$ is the critical density.
Note that we have  only three parameters, 
 $\alpha_{B-L} = g_{B-L}^2/(4\pi)$, $m_{Z'}$ and $m_{\textrm{DM}}$, 
 in our analysis. 
For $m_{Z'} = 3$ TeV and various values of the gauge coupling $\alpha_{B-L}$,
Figure \ref{BLomega} depicts  the resultant dark matter relic density as a function of its mass $m_{\textrm{DM}}$,
  along with the observed bounds $0.1183 \le \Omega_{\textrm{DM}} h^2 \le 0.1213$ \cite{Aghanim:2015xee} (two horizontal dashed lines).
The solid curves from top to bottom correspond to the results for $\alpha_{B-L} = 0.001, \ 0.0014, \ 0.002, \ 0.003$ and $0.005$,
respectively.
We find that in order to reproduce the observed relic density, 
  the dark matter mass must be close to half of the $Z_{B-L}^{\prime}$ boson mass. 
In other words, normal values of the dark matter annihilation cross section leads to overabundance,
and it is necessary that an enhancement of the cross section through the $Z_{B-L}^{\prime}$ boson resonance
in the $s$-channel annihilation process.
In Figure \ref{BLomega}, we can see the maximum annihilation cross section occurs 
 for $m_{\textrm{DM}}$ slightly smaller than $m_{Z^\prime}/2$ 
 because of the effect from thermally averaging the cross section.

%
%

%
\begin{figure}[tb]
 \begin{center}
  \includegraphics[width=80mm]{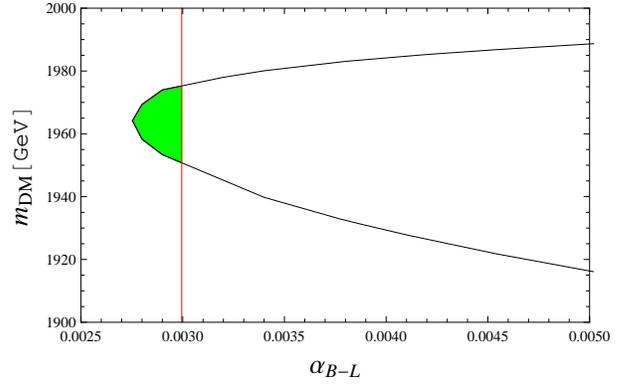}
 \end{center}
 \renewcommand{\baselinestretch}{1.1}
 \caption{ 
 The dark matter mass as a function of $\alpha_{B-L}$ for $m_{Z'} = 4$ TeV.
 Along the solid (black) curve in each panel, $\Omega_{\textrm{DM}} h^2 = 0.1198$ is satisfied.
 The vertical solid line (in red) represents the upper bound on $\alpha_{B-L}$ obtained from
 the ATLAS results \cite{ATLAS:2017} (see Figures \ref{BLfinal}).
 The (green) shaded region satisfies $\Omega_{\textrm{DM}} h^2 \leq 0.1198$ 
 and the ATLAS bound on $\alpha_{B-L} \leq 0.003$.  
} 
\label{BLAvsM34}
\end{figure}
%

As can be seen (\ref{annihilation_X}), the dark matter annihilation cross section becomes smaller 
  as the gauge coupling $\alpha_{B-L}$ is lowered, for a fixed $m_{\textrm{DM}}$. 
This can be seen in Figure \ref{BLomega}, where for a fixed $m_{\textrm{DM}}$, 
  the resultant relic abundance becomes larger as $\alpha_{B-L}$ is lowered.  
As a result, there is a lower bound on $\alpha_{B-L}$ in order to satisfy the cosmological upper bound
   on the dark matter relic abundance $\Omega_{\textrm{DM}} h^2 \le 0.1213$.
For a $\alpha_{B-L}$ value larger than the lower bound ($\alpha_{B-L}=0.0014$ in Figure \ref{BLomega}), we can find two values of $m_{DM}$
   which result in the center value of the observed relic abundance $\Omega_{\textrm{DM}} h^2 = 0.1198$.
In Figure \ref{BLAvsM34}, we show the dark matter mass yielding $\Omega_{\textrm{DM}} h^2 = 0.1198$
 as a function of $\alpha_{B-L}$, for $m_{Z'} = 4$ TeV. 
As a reference, we also show the dotted lines corresponding to $m_{\textrm{DM}} = m_{Z'}/2$.
In Figure \ref{BLomega}, we see that the minimum relic abundance is achieved
   by a dark matter mass which is very close to, but smaller than $m_{Z'}/2$.
Although the annihilation cross section of (\ref{xsection}) has a peak at $\sqrt{s} = m_{Z'}$,
   the thermal averaged cross section given in (\ref{avesv}) includes the integral of the product of
   the reduced cross section and the modified Bessel function $K_1$.
Our results indicate that for $m_{\textrm{DM}}$ taken to be slightly smaller than $m_{Z'}/2$,
   the thermal averaged cross section is larger than the one for $m_{\textrm{DM}} = m_{Z'}/2$.

\begin{figure}[tb]
 \begin{center}
  \includegraphics[width=80mm]{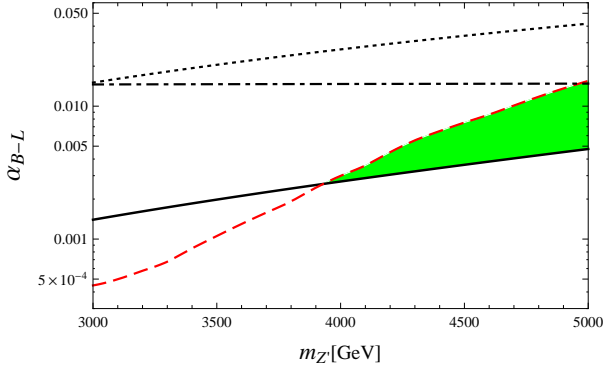}
 \end{center}
 \renewcommand{\baselinestretch}{1.1}
 \caption{
Allowed parameter region for the $Z'_{B-L}$ portal dark matter scenario.
The solid (black) line shows the lower bound on $\alpha_{B-L}$ as a function of $m_{Z'}$ to satisfy
  the cosmological upper bound on the dark matter relic abundance.
The dashed line (in red) shows the upper bound on $\alpha_{B-L}$ as a function of $m_{Z'}$ from the search results
 for $Z'$ boson resonance by the ATLAS collaboration \cite{ATLAS:2017}.
The LEP bound is depicted as the dotted line.
Combining these bounds, the allowed parameter region is depicted as the (green) shaded region. 
We also show a theoretical upper bound on $\alpha_{B-L}$ (dashed-dotted) to avoid the Landau pole of the running $B-L$ gauge coupling
  below the Planck mass $M_{pl}$.
} 
 \label{BLfinal}
\end{figure}

As mentioned above, for a fixed $Z'_{B-L}$ boson mass,
   we can find a corresponding lower bound on the gauge coupling $\alpha_{B-L}$ in order for the resultant
   relic abundance not to exceed the cosmological upper bound $\Omega_{\textrm{DM}} h^2 = 0.1213$.
Figure \ref{BLfinal} depicts the lower bound of $\alpha_{B-L}$ as a function of $m_{Z'}$ [solid (black) line].
Along this solid (black) line, we find that the dark matter mass is approximately given by $m_{\textrm{DM}} \simeq 0.49 m_{Z'}$.
The dark matter relic abundance exceeds the cosmological upper bound in the region below the solid (black) line.
Along with the other constraints that will be obtained in the next section, Figure \ref{BLfinal} is our main results in this section.

%
%

\section{LHC Run-2 constraints}
\label{sec:5}

%
\begin{figure}
 \begin{center}
  \includegraphics[width=80mm]{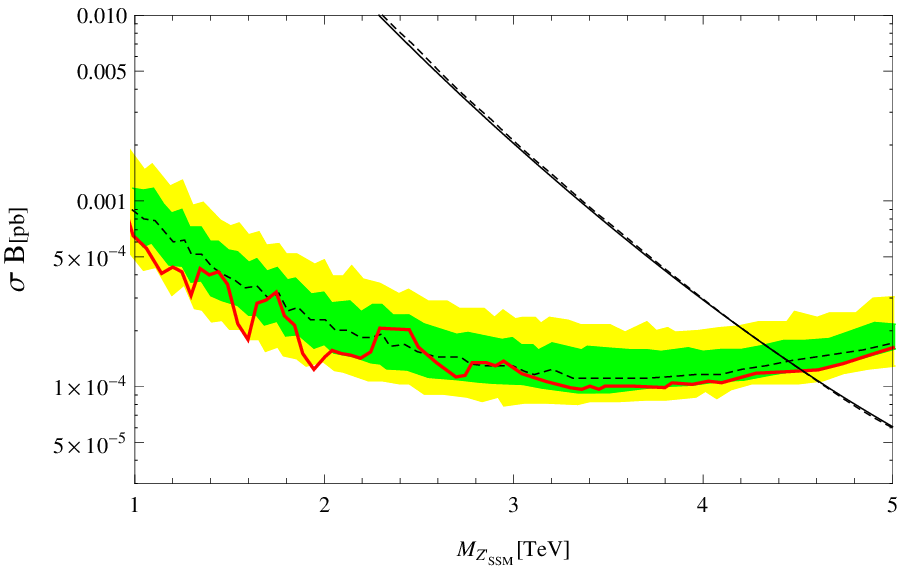}
  \includegraphics[width=80mm]{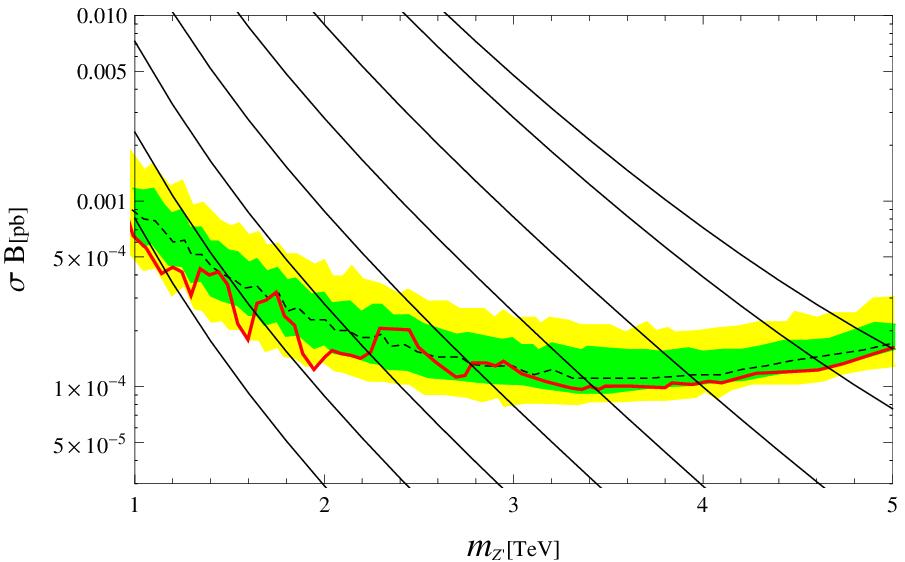}
 \end{center}
 \renewcommand{\baselinestretch}{1.1}
 \caption{Top panel: the cross section as a function of the $Z'_{SSM}$ mass (solid line) with $k=1.31$,
 along with the ATLAS result in Ref.~\cite{ATLAS:2017} from the combined dielectron and dimuon channels.
 Bottom panel: the cross sections calculated for various values of $\alpha_{B-L}$ with $k = 1.31$.
 The solid lines from left to right correspond to 
   $\alpha_{B-L} = 10^{-5}, \ 10^{-4.5}, \ 10^{-4}, \ 10^{-3.5}, \ 10^{-3}, \ 10^{-2.5}, \ 10^{-2}$, and $10^{-1.8}$, respectively.
 } 
\label{BLATL13TeV}
\end{figure}
%


The ATLAS and the CMS collaborations have been searching for a $Z'$ boson resonance with dilepton final states\footnote{
Although the $Z'$ boson resonance has been searched also with dijet final states \cite{Aaboud:2017yvp}, 
  we can see that the constraints from this search is weaker than the one with dilepton final states. 
}
   at the LHC Run-2 \cite{ATLAS:2016, CMS:2016} (for the process, see the right diagram in Figure \ref{BLcomplementarity}),
   and have improved the upper limits of the $Z'$ boson production cross section from those in the LHC Run-1 \cite{ATLAS8TeV, CMS8TeV}.
Employing the LHC Run-2 results, in particular, the most recent ATLAS result 
  with a $36/$fb luminosity\cite{ATLAS:2017}, 
   we will derive an upper bound on $\alpha_{B-L}$ as a function of $m_{Z'}$.\footnote{
The results in this article are the update of the results in Refs.~\cite{Okada:2016gsh, Okada:2016tci}. 
}  
Since we have obtained in the previous section the lower bound on $\alpha_{B-L}$
   as a function of $m_{Z'}$ from the constraint on the dark matter relic abundance,
   the LHC Run-2 results are complementary to the cosmological constraint.
As a result, the parameter space of the $Z'_{B-L}$ portal dark matter scenario
   is severally constrained once the two constraints are combined.

Let us consider the $Z'_{B-L}$ boson production process,  $pp \to Z'_{B-L} + X \to l^+ l^- + X$, 
 where $X$ denotes hadron jets. 
The differential cross section is given by 
\begin{eqnarray} \label{dxsection}
 \frac{d\sigma}{dM_{ll}}
 = &&
 \sum_{a,b} \int^1_{\frac{M_{ll}^2}{E_{\textrm{CM}}^2}} dx \frac{2M_{ll}}{x E_{\textrm{CM}}^2}
 f_a(x, Q^2) f_b \left( \frac{M_{ll}^2}{E_{\textrm{CM}}^2}, Q^2 \right) \nonumber \\
 && \times \hat{\sigma}( q\bar{q} \to  Z'_{B-L} \to l^+ l^-),
\end{eqnarray}
where  $E_{\textrm{CM}} = 13$ TeV is the LHC Run-2 energy in the center-of-mass frame, 
  $M_{ll}$ is the invariant mass of the dilepton final state,  
  and $f_a$ is the parton distribution function (PDF) for a parton “$a$.” 
For the PDFs we utilize CTEQ6L \cite{Pumplin:2002vw} with $Q = m_{Z'}$ as the factorization scale. 
Here, the cross section for the colliding partons is given by
\begin{eqnarray}
 \hat{\sigma}
 = \frac{4\pi \alpha_{B-L}^2}{81} \frac{M_{ll}^2}{( M_{ll}^2 - m_{Z'}^2 )^2 + m_{Z'}^2 \Gamma_{Z'}^2}.
\end{eqnarray}
In calculating the total cross section, we set a range of $M_{ll}$ that is used 
  in the analysis by the ATLAS and the CMS collaborations, respectively. 
We compare our results of the total cross section with the upper limits of the ATLAS and CMS results.

In the analysis by the ATLAS and the CMS collaborations,
   the so-called sequential SM $Z'$ ($Z'_{SSM}$) model \cite{Barger:1980ti} has been considered as a reference model.
We first analyze the sequential $Z'$ model to check a consistency of our analysis with the one by the ATLAS collaboration \cite{ATLAS:2017}. 
In the sequential $Z'$ model, the $Z'_{SSM}$ boson has exactly the same couplings with quarks and leptons as the SM Z boson.
With the couplings, we calculate the cross section of the process $pp \to Z'_{SSM} + X \to l^+ l^- + X$ like (\ref{dxsection}).
By integrating the differential cross section in the region of 128 GeV $\le M_{ll} \le$ 6000 GeV \cite{ATLAS8TeV},
   we obtain the cross section of the dilepton production process as a function of $Z'_{SSM}$ boson mass.
Our result is shown as a solid line in the top panel on Figure \ref{BLATL13TeV}, along with the plot presented
   by the ATLAS collaboration \cite{ATLAS:2017}.
In the analysis in the ATLAS paper, the lower limit of the $Z'_{SSM}$ boson mass is found to be $4.5$ TeV,
   which is read from the intersection point of the theory prediction (diagonal dashed line)
   and the experimental cross section bound [horizontal solid curve (in red)].
In order to take into account the difference of the PDFs used in the ATLAS
   and our analysis and QCD corrections of the process,
   we have scaled our resultant cross section by a factor $k = 1.31$,
   with which we can obtain the same lower limit of the $Z'_{SSM}$ boson mass as $4.5$ TeV.
We can see that our result with the factor of  $k = 1.31$ is very consistent
   with the theoretical prediction (diagonal dashed line) presented in Ref.~\cite{ATLAS:2017}.
This factor is used in our analysis of the $Z'_{B-L}$ production process.
Now we calculate the cross section of the process $pp \to Z'_{B-L} + X \to l^+ l^- + X$ for various values of $\alpha_{B-L}$,
   and our results are shown in the bottom panel of Figure \ref{BLATL13TeV}, along with the plot in Ref.~\cite{ATLAS:2017}.
The diagonal solid lines from left to right correspond to 
  $\alpha_{B-L} = 10^{-5}, \ 10^{-4.5}, \ 10^{-4}, \ 10^{-3.5}, \ 10^{-3}, \ 10^{-2.5}, \ 10^{-2}$, and $10^{-1.8}$, respectively.
From the intersections of the horizontal curve and diagonal solid lines,
   we can read off a lower bound on the $Z'_{B-L}$ boson mass for a fixed $\alpha_{B-L}$ value.
In this way, we have obtained the upper bound on $\alpha_{B-L}$ as a function the $Z'_{B-L}$ boson mass,
   which is depicted in Figure \ref{BLfinal} [dashed (red) line].

In Figure \ref{BLfinal}, we also show the LEP bound as the dotted line which is obtained from the search
   for effective 4-Fermi interactions mediated by the $Z'_{B-L}$ boson \cite{Carena:2004xs}. 
An updated limit with the final LEP 2 data \cite{LEP:2013} is found to be \cite{Heeck:2014zfa}
\begin{eqnarray}
 \frac{m_{Z'}}{g_{B-L}} \ge 6.9 \ {\rm{TeV}}
\end{eqnarray}
   at 95\% confidence level.
We find that the ATLAS bound at the LHC Run-2 is more severe than the LEP bound 
   for the $Z'$ boson mass range presented here. 
In order to avoid the Landau pole of the running $B-L$ coupling $\alpha_{B-L}(\mu)$,
   below the Plank mass ($1/\alpha_{B-L}(M_{pl}) > 0$), we find
\begin{eqnarray}
 \alpha_{B-L} < \frac{\pi}{6 \ln{ \left[ \frac{M_{pl}}{m_{Z'}} \right]}},
\end{eqnarray}
   which is shown as the dashed-dotted line in Figure \ref{BLfinal}.
Here, the gauge coupling $\alpha_{B-L}$ used in our analysis for dark matter physics and LHC physics
   is defined as the running gauge coupling $\alpha_{B-L}(\mu)$ at $\mu = m_{Z'}$,
   and we have employed the renormalization group equation at the one-loop level with $m_N^1=m_N^2=m_{\Phi}=m_{Z'}$, for simplicity.

\section{Conclusions}
\label{sec:6}

We have discussed a simple extension of the SM where the global $B-L$ symmetry in the SM 
  is promoted to the $B-L$ gauge symmetry. 
In the minimal version of this extension, which is the so-called minimal $B-L$ model, 
  we introduce three right-handed neutrinos with a $B-L$ charge $-1$ and 
  the $B-L$ (SM singlet) Higgs field with a $B-L$ charge $+2$. 
The three right-handed neutrinos cancel all the gauge and gravitational anomalies 
  caused by gauging the $B-L$ symmetry.
The VEV of  the $B-L$ Higgs field breaks the $B-L$ gauge symmetry 
  and generates the $B-L$ gauge boson ($Z^\prime_{B-L}$) mass but also the Majorana masses 
  for the right-handed neutrinos.  
The SM neutrino mass matrix is then generated after the electroweak symmetry breaking. 
In order to supplement the minimal $B-L$ model with a dark matter candidate, 
  we have introduced a Z$_2$ symmetry and one right-handed neutrino of a unique $Z_2$-odd particle in the model 
  plays the role of the dark matter. 
In this way, the minimal $B-L$ model with $Z_2$ symmetry supplements 
  the major missing pieces of the SM, the neutrino mass matrix and a dark matter candidate, 
  while the original particle content of the minimal $B-L$ model is kept intact. 

In this model context, we have investigated the “$Z_{B-L}^{\prime}$ portal” dark matter scenario,  
  where the dark matter particle ($Z_2$-odd right-handed neutrino) mainly communicates with 
  the SM particles through the $Z_{B-L}^{\prime}$ boson.  
We have only three free parameters in our analysis, namely, 
  the gauge coupling ($\alpha_{B-L}$), the dark matter mass ($m_{\textrm{DM}}$), 
  and the $Z_{B-L}^{\prime}$ boson mass ($m_{Z^{\prime}}$).
We have derived the lower bound on $\alpha_{B-L}$ as a function of $m_{Z^{\prime}}$ 
   by using the cosmological bound on the dark matter relic abundance. 
On the other hand, the LHC Run-2 results on the search for a narrow resonance 
  constrain the $Z'_{B-L}$ production cross section at the LHC. 
We have interpreted the latest results by the ATLAS collaboration \cite{ATLAS:2017} 
  and derived the upper bound on $\alpha_{B-L}$ as a function of $m_{Z^{\prime}}$. 
Similar (but weaker) upper bounds on $\alpha_{B-L}$ have been obtained from the results 
  by the LEP experiment and the perturbativity condition of the running $B-L$ gauge coupling 
  below the Planck mass.
After combining all constraints, we have obtained the allowed parameter space shown in Figure \ref{BLfinal}.
We can see that the cosmological and the collider constraints are complementary for narrowing down 
   the arrowed parameter space:  $m_{Z^{\prime}} \ge 3.9$ TeV.
Since the SM background events for $m_{Z^{\prime}} \gtrsim3$ TeV are negligibly small (see Ref.~\cite{ATLAS:2017}), 
   we expect that the future search reach for the $Z^\prime$ boson production scales as the luminosity of the LHC experiments. 
In the narrow decay width approximation (which is justified in our analysis), the $Z^\prime_{B-L}$ boson production cross section 
   is proportional to $\alpha_{B-L}$. 
Therefore, from Figure \ref{BLfinal}, we expect that the (green) shaded region will be covered 
   in the future with a LHC luminosity about 120/fb.

Towards direct and indirect detections of dark matter particles, many experiments are in operation and also planned. 
Because of its Majorana nature, the right-handed neutrino dark matter has the axial vector coupling 
  with the $Z^\prime_{B-L}$ boson, while the SM fermions have the vector couplings due to their $B-L$ charges. 
Hence, the elastic scattering amplitude between the dark matter particles and quarks 
   through the $Z^\prime_{B-L}$ boson exchange 
   is vanishing in the non-relativistic limit, and our dark matter particle evades its direct detection. 
We can consider an indirect detection of the right-handed neutrino dark matter 
  through cosmic rays from their pair annihilations in the galactic halo.  
However, because of the Majorana nature, the pair annihilation cross section is highly suppressed 
  by a dark matter velocity at the present universe, and we find that the cosmic ray flux from the dark matter 
  pair annihilations is far below the observable level for the parameter region shown in Figure \ref{BLfinal}.

If kinematically allowed, a pair of right-handed neutrinos involved in the seesaw mechanism
  can be produced from a $Z^\prime_{B-L}$ boson decay. 
Collider signatures of the right-handed neutrinos produced in this way have been studied 
  (see, for example, Refs.~\cite{Kang:2015uoc, Cox:2017eme, Accomando:2017qcs, Das:2017flq, Das:2017deo} for recent studies). 
The right-handed neutrinos, once observed at the LHC, are clue to understand the mass generation mechanism 
  of light neutrinos. 
In addition to the study of $Z^\prime$ boson production at the LHC, 
   the search for right-handed neutrinos at the future LHC is worth investigating.

It is interesting to extend the minimal $B-L$ model to the so-called non-exotic $U(1)_X$ model \cite{Appelquist:2002mw}.  
In this model, the particle content remains the same, 
  while the $U(1)_X$ charge of a particle is generalized as $Q_X= Y x_H + Q_{B-L}$, 
  where $Y$ and $Q_{B-L}$ are $U(1)_Y$ and $U(1)_{B-L}$ charges of the particle, respectively,  
  and $x_H$ is a real parameter. 
In this $U(1)_X$ generalization, the minimal $B-L$ model is realized as a limit $x_H=0$. 
For studies on the $Z^\prime$ portal dark matter scenario in the minimal $U(1)_X$ model, see Ref.~\cite{Okada:2016tci}. 
With a special value of $x_H=-4/5$, we can consider the unification of the model 
  into the gauge group $SU(5) \times U(1)_X$ \cite{Okada:2017dqs}.

Finally, our minimal $B-L$ model with the right-handed neutrino dark matter 
   can also account for the origin of the baryon asymmetry in the universe through leptogenesis \cite{Fukugita:1986hr}  
   with two Z$_2$-even right-handed neutrinos if they are almost degenerate (so-called resonant leptogenesis \cite{Pilaftsis:1997jf, Pilaftsis:2003gt}).  
See Ref.~\cite{Iso:2010mv} for detailed analysis. 
Furthermore, if we introduce non-minimal gravitational coupling, 
   the $B-L$ Higgs field plays the role of inflaton which causes cosmological inflation in the early universe.  
We can achieve the successful cosmological inflation scenario with a suitable choice of the non-minimal gravitational coupling constant. 
See, for example, Refs.~\cite{Okada:2011en, Okada:2015lia, Oda:2017zul}.

\section*{Acknowledgments}

The author would like to appreciate Nobuchika Okada for collaborations.

\addcontentsline{toc}{chapter}{References}
\bibliographystyle{utphysII} 
%
{
\def\chapter*#1{}
\bibliography{./bibliography}                           
}

\end{document}